\documentclass{jps-cp}
\usepackage{txfonts}
\usepackage{graphicx,amsmath,amssymb,wrapfig}
\usepackage{slashed}

\title{TMDs for spin-1 hadrons}
\author{S. \textsc{Kumano}$^{1,2}$ and Qin-Tao \textsc{Song}$^{3,4}$}
\inst{$^{1}$ KEK Theory Center,
             IPNS, 
             KEK,
             Oho 1-1, Tsukuba, Ibaraki, 305-0801, Japan\\
      $^{2}$ J-PARC Branch, KEK Theory Center,
             IPNS, KEK,
             and Theory Group, Particle and Nuclear Physics Division, 
             J-PARC Center, 
             Shirakata 203-1, Tokai, Ibaraki, 319-1106, Japan\\
      $^{3}$ School of Physics and Microelectronics, Zhengzhou University, 
             Zhengzhou, Henan 450001, China\\
      $^{4}$ CPHT, CNRS, Ecole Polytechnique, Institut Polytechnique de Paris,
             Route de Saclay, 91128 Palaiseau, France}
\email{shunzo.kumano@kek.jp,songqintao@zzu.edu.cn}

\recdate{January 14, 2022}

\abst{
Transverse-momentum-dependent parton distribution functions (TMDs)
were investigated at the twists 3 and 4 for spin-1 hadrons
in addition to the twist-2 ones. They were found by studying 
all the possible  decomposition of a quark correlation function
in a Lorentz-invariant way with the Hermiticity and parity invariance.
The time-reversal invariance was not imposed for the TMDs
due to an active role of gauge links; however, they were
used for collinear parton distribution functions (PDFs)
by integrating the TMDs over the transverse momentum.
We found that 30 TMDs exist in the tensor-polarized spin-1 hadron
at the twists 3 and 4, whereas there are 10 TMDs at the twist 2.
We also showed that there are 3 collinear PDFs at the twists 3 and 4.
The corresponding TMD fragmentation functions exist 
at the twists 2, 3, and 4 simply by changing function names and variables.
Since the time-reversal invariance is valid in the collinear PDFs,
the integrals of time-reversal-odd TMDs 
over the transverse momentum should vanish. 
It leads to the sum rules
$ \int d^2 k_T \, h_{1LT} (x, k_T^{\, 2}) 
= \int d^2 k_T g_{LT} (x, k_T^{\, 2}) 
= \int d^2 k_T h_{LL} (x, k_T^{\, 2}) 
= \int d^2 k_T h_{3LL}(x, k_T^{\, 2}) =0$
on the time-reversal odd TMDs at the twists 3 and 4.
}

\kword{QCD, quark, gluon, spin-1 hadron, TMD, higer twist}

\begin{document}
\maketitle

\vspace{-1.00cm}
\section{Introduction}
\vspace{-0.20cm}

In the field of nucleon structure, recent interests tend to focus on
three-dimensional structure functions, namely 
generalized parton distributions (GPDs),
generalized distribution amplitudes (GDAs or timelike GPDs),
and transverse-momentum-dependent parton distribution functions (TMDs).
These functions play a crucial role in determining the origins
of hadron spins and masses.
Among them, the TMD physics is unique for finding explicit color
degrees of freedom in terms of color flow within hadrons.
It is an interesting interdisciplinary physics field,
for example, in connection with the gluon condensate,
the color Aharonov-Bohm effect, and the color entanglement.

On the other hand, structure functions of spin-1 hadrons could become
an interesting field of high-energy spin physics because of additional
functions to the ones of the spin-1/2 nucleons.
The spin-1 deuteron has been described 
as a simple bound system of a proton and a neutron
in traditional nuclear physics. However, new hadron-physics
aspects could be probed by tensor-polarized structure functions,
such as $b_1$ \cite{b1}
and the gluon transversity \cite{g-transversity-drell-yan}.
Such aspects will be investigated experimentally in the near future
at various accelerator facilities, such as 
the Thomas Jefferson National Accelerator Facility (JLab), 
Fermilab (Fermi National Accelerator Laboratory), 
Nuclotron-based Ion Collider fAcility (NICA),
LHC (Large Hadron Collider)-spin,
and electron-ion colliders (EIC, EicC).
Considering this situation, we investigated possible TMDs,
collinear parton distribution functions (PDFs),
and fragmentation functions up to the twist 4
for spin-1 hadrons \cite{ks-tmd-2021}, 
so that the spin-1 TMDs, the PDFs, and the fragmentation functions
can be studied at the same level with the spin-1/2 structure functions
by including higher-twist effects.
In this report, we explain our recent results.

\section{Decomposition of a quark correlation function
         in a tensor-polarized spin-1 hadron}
\vspace{-0.10cm}

The TMDs and PDFs are defined from a correlation function
\vspace{-0.10cm}
\begin{align}
\Phi_{ij}^{[c]} (k, P, T  \, | \, n )
& =
\int  \! \frac{d^4 \xi}{(2\pi)^4} \, e^{ i k \cdot \xi}
\langle \, P , T \left | \, 
\bar\psi _j (0) \,  W^{[c]} (0, \xi)  
 \psi _i (\xi)  \, \right | P, \,  T \, \rangle ,
\label{eqn:correlation-q}
\\[-0.80cm] \nonumber
\end{align} 
which is the amplitude to extract a parton from a hadron and 
then to insert it into the hadron at a different space-time point.
Since only the difference from the spin-1/2 nucleon case is
the tensor-polarization part, we restrict our studies to
the tensor polarization $T^{\mu\nu}$ described by the polarization parameters 
($S_{LL}$, $S_{LT}^\mu$, $S_{TT}^{\mu\nu}$) as
\vspace{-0.10cm}
\begin{align}
T^{\mu\nu}  = \frac{1}{2} &  \left [ \frac{4}{3} S_{LL} \frac{(P^+)^2}{M^2} 
               \bar n^\mu \bar n^\nu 
          - \frac{2}{3} S_{LL} ( \bar n^{\{ \mu} n^{\nu \}} -g_T^{\mu\nu} )
\right.
\nonumber \\[-0.10cm]
& 
\left.
+ \frac{1}{3} S_{LL} \frac{M^2}{(P^+)^2}n^\mu n^\nu
+ \frac{P^+}{M} \bar n^{\{ \mu} S_{LT}^{\nu \}}
- \frac{M}{2 P^+} n^{\{ \mu} S_{LT}^{\nu \}}
+ S_{TT}^{\mu\nu} \right ].
\label{eqn:spin-1-tensor-1}
\\[-0.80cm] \nonumber
\end{align}
In these equations, $k$ and $P$ are quark and hadron momenta, 
$\psi$ is the quark field, $\xi$ is the space-time coordinate,
$ W^{[c]} (0, \xi)$ is the gauge link with the integral path $c$,
$n$ and $\bar n$ are the lightcone vectors 
$n^\mu =(1,0,0,-1)/\sqrt{2}$ and $\bar n^\mu =(1,0,0,1)/\sqrt{2}$,
$g_T^{\mu\nu}$ is given by $g_T^{11}=g_T^{22}=-1$ and the others$\ =0$,
$a^{\{ \mu} b^{\nu \}}$ indicates the symmetrized combination
$a^{\{ \mu} b^{\nu \}} = a^\mu b^\nu + a^\nu b^\mu$,
$M$ is the hadron mass, 
and $P^+$ is the lightcone momentum given by $P^+ =(P^0 +P^3)/\sqrt{2}$.
In Eq.\,(\ref{eqn:correlation-q}), the vector polarization $S^\mu$ is
not explicitly written because only the tensor
polarization is investigated in this paper.

Taking into account the constraints of the Hermiticity and parity invariance,
we decompose the quark correlation function in a Lorentz-invariant way as
\vspace{-0.10cm}
\begin{align}
\Phi(k, P, T \, | n) & = \frac{A_{13}}{M}  T_{kk} + \frac{A_{14}}{M^2} T_{kk} 
     \slashed{P}
+ \cdots
+ \frac{A_{20}}{M^2} \varepsilon^{\mu\nu P k}  \gamma_{\mu} \gamma_5 T_{\nu k}
\nonumber \\[-0.10cm]
&
+ \frac{B_{21}M}{P\cdot n} T_{kn}  +\frac{B_{22}M^3}{(P\cdot n)^2} T_{nn}
+ \cdots
+ \frac{B_{52}M}{P\cdot n } \sigma_{\mu k}  T^{\mu n} .
\label{eqn:cork4}
\\[-0.80cm] \nonumber
\end{align} 
Its full expression is given in Eq.\,(20) of Ref.\,\cite{ks-tmd-2021}.
Here, $A_i$ and $B_i$ are expansion coefficients, and 
the contraction $X_{\mu k} \equiv X_{\mu \nu} k^{\nu}$ is used.
This expansion is an extension of the work in Ref.\,\cite{bm-2000}
by including the additional $n$ terms so as to find
twist 3 and twist 4 functions, as it was done 
in the spin-1/2 nucleons \cite{tmds-nucleon}.
The TMD and collinear correlation functions are given by
integrating over the quark momenta as
\vspace{-0.30cm}
\begin{align}
\Phi^{[c]} (x, k_T, P, T ) & = \! \int \! dk^+ dk^- \, 
               \Phi^{[c]} (k, P, T  \, |n ) \, \delta (k^+ \! -x P^+) ,
\label{eqn:correlation-tmd}
\\[-0.05cm]
\Phi (x, P, T ) & 
  = \! \int \! d^2 k_T \, \Phi^{[c]} (x, k_T, P, T ) .
\label{eqn:correlation-pdf}
\\[-0.70cm] \nonumber
\end{align}

\section{TMDs and PDFs for a spin-1 hadron up to twist 4}
\vspace{-0.10cm}

The TMDs and collinear PDFs are defined by taking traces of 
the TMD and collinear correlation functions in 
Eqs.\,(\ref{eqn:correlation-tmd}) and (\ref{eqn:correlation-pdf})
with various $\gamma$ matrices ($\Gamma$) as
$ \Phi^{\left[ \Gamma \right]} \equiv 
\frac{1}{2} \, \text{Tr} \left[ \, 
\Phi \Gamma \, \right] $.
The twist-2 functions were studied 
by calculating $\Phi^{ [ \gamma^+ ] }$,
$\Phi^{ [ \gamma^+ \gamma_5 ] }$, and
$\Phi^{ [ i \sigma^{i+} \gamma_5 ] }$
(or $\Phi^{ [ \sigma^{i+} ] }$) \cite{bm-2000}.
The twist-2 TMDs and PDFs are listed in 
Tables \ref{table:twist-2-tmds} and \ref{table:twist-2-pdfs}.
The twist-3 and 4 results are our studies \cite{ks-tmd-2021}
in Tables 
\ref{table:twist-3-tmds}, \ref{table:twist-3-pdfs},
\ref{table:twist-4-tmds}, and \ref{table:twist-4-pdfs}.
The twist-3 TMDs were obtained by taking the traces,
$\Phi^{ [ \gamma^i ] }$,
$\Phi^{\left[{\bf 1}\right]}$,
$\Phi^{\left[i\gamma_5\right]}$
$\Phi^{ [\gamma^{i}\gamma_5 ]}$
$\Phi^{ [ \sigma^{ij} ]}$,
and $\Phi^{ [ \sigma^{-+} ] }$.
The $i$ and $j$ are transverse indices ($i,j=1$ or 2).
The twist-4 TMDs were obtained by
$\Phi^{[\gamma^-]}$,
$\Phi^{[\gamma^- \gamma_5]}$, and $\Phi^{[\sigma^{i-}]}$.
For example, some of twist-3 TMDs are defined in the trace 
$\Phi^{ [ \gamma^i ] }$ as
\vspace{-0.20cm}
\begin{align}
\Phi^{ [ \gamma^i ] } (x, k_T, T)
= 
\frac{M}{P^+} \bigg [ & f^{\perp}_{LL}(x, k_T^{\, 2})  S_{LL} \frac{k_T^i}{M}
\! + \! f^{\,\prime} _{LT} (x, k_T^{\, 2})S_{LT}^i 
- f_{LT}^{\perp}(x, k_T^{\, 2}) \frac{ k_{T}^i  S_{LT}\cdot k_{T}}{M^2} 
\nonumber \\[-0.10cm]
& 
- f_{TT}^{\,\prime} (x, k_T^{\, 2}) \frac{S_{TT}^{ i j} k_{T \, j} }{M} 
+ f_{TT}^{\perp}(x, k_T^{\, 2}) \frac{k_T\cdot S_{TT}\cdot k_T}{M^2} 
       \frac{k_T^i}{M} \bigg ] .
\label{eqn:cork-3-1a}
\nonumber \\[-0.70cm]
\end{align} 
The TMDs with the prime ($^\prime$) are redefined as
$ F (x, k_T^{\, 2}) \equiv F^{\,\prime} (x, k_T^{\, 2})
 - (k_T^{\, 2} /(2M^2)) \, F^{\perp} (x, k^{\, 2}_T) $
with $k_T^{\, 2}= - \vec k_T^{\, 2}$, so that
the functions without the primes are shown in the tables.

In the tables, the polarizations U, L, and T indicate
unpolarized, longitudinally polarized,
and transversely polarized, respectively,
and they also exist in the spin-1/2 nucleons.
The tensor polarizations (LL, LT, TT) are additional 
in the spin-1 hadrons.
The time-reversal even (T-even) and odd (T-odd) distributions 
are classified in the tables.
Chiral-odd distributions are shown with the square brackets $[\ ]$,
and the distributions without the bracket are chiral-even ones. 
The asterisks $*1$,\,$*2$,\,$*3$,\,$*4$ indicate that
the collinear PDFs
$h_{1LT} (x)$,\,$g_{LT} (x)$,\,$h_{LL} (x)$,\,$h_{3LT} (x)$
vanish, respectively, due to the time-reversal invariance;
however, the corresponding fragmentation functions
$H_{1LT} (z)$,\,$G_{LT} (z)$,\,$H_{LL} (z)$,\,$H_{3LT} (z)$
should exist as collinear fragmentation functions
\cite{ks-tmd-2021,ji-ffs}.
Furthermore, finite transverse-momentum moments could exist
even for the T-odd TMDs \cite{eq-motion}.
In the tensor-polarized spin-1 hadron,
there are 40 TMDs in total,
and there are 20 TMDs at the twist 3 and 10 TMDs at the twist 4.
There are four PDFs in total, 
and there are two at the twist 3 and one at the twist 4
in the tensor polarizations.
Since the time-reversal invariance is valid in the collinear PDFs,
we have the sum rules for the T-odd TMDs as
\vspace{-0.25cm}
\begin{align}
\! \int \! d^2 k_T \, h_{1LT} (x, k_T^{\, 2}) 
= \! \int \! d^2 k_T \, g_{LT} (x, k_T^{\, 2}) 
= \! \int \! d^2 k_T \, h_{LL} (x, k_T^{\, 2}) 
= \! \int \! d^2 k_T \, h_{3LT}(x, k_T^{\, 2})  = 0 .
\label{eqn:TMD-sum}
\\[-0.85cm]
\nonumber 
\end{align} 
There are an additional sum rule and useful relations
in these PDFs and multiparton distribution functions
\cite{eq-motion,ks-ww-bc-2021}.
Now, it became possible to investigate structure functions
of spin-1 hadrons up to twist 4 in the same way with
the spin-1/2 nucleons.

\begin{table}[h!]
\vspace{-0.55cm}
\begin{minipage}{0.45\textwidth}
  \includegraphics[width=6.39cm]{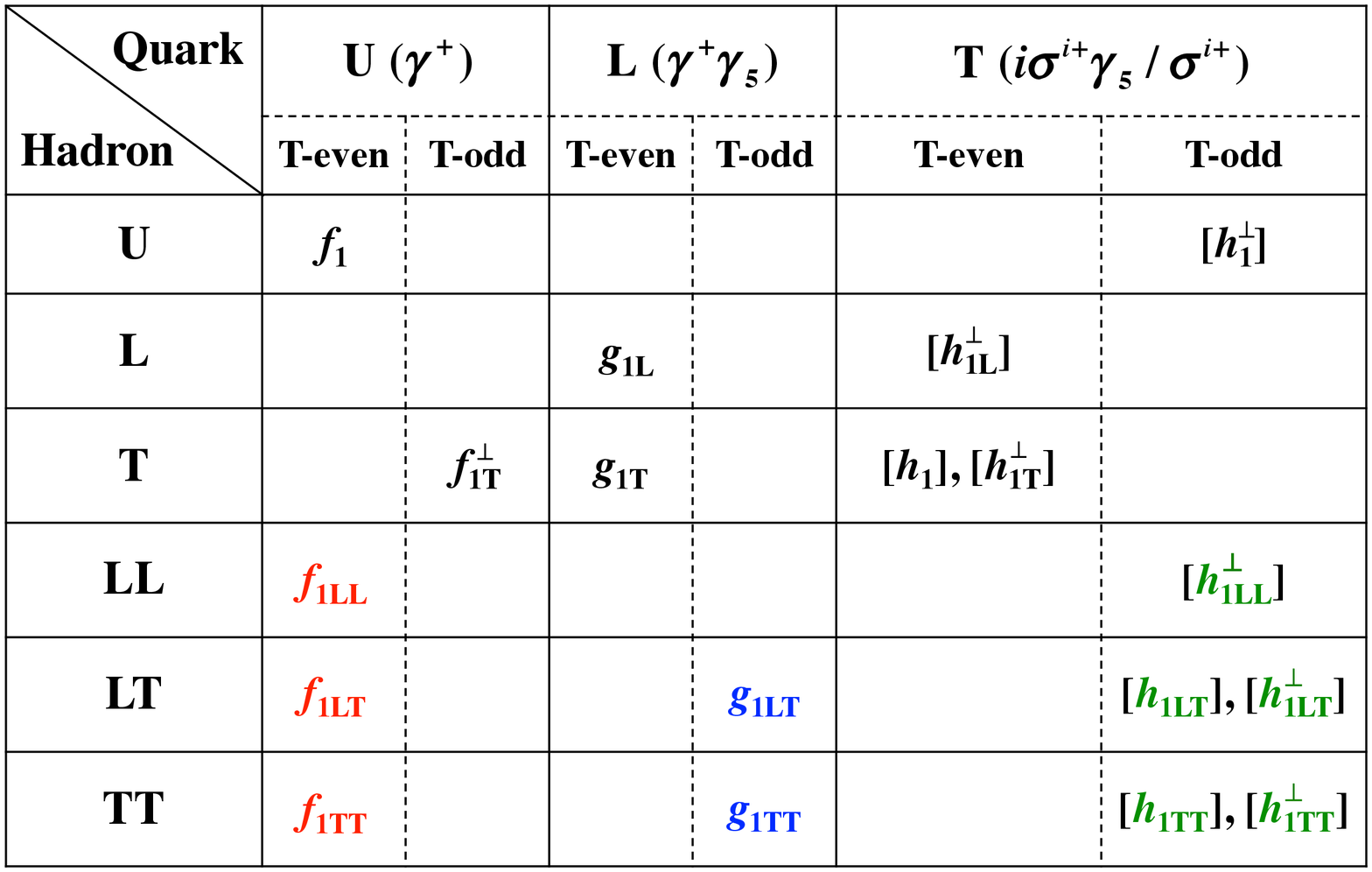}
  \vspace{-0.15cm}
  \caption{\hspace{-0.10cm} Twist-2 TMDs.}
  \label{table:twist-2-tmds}
\end{minipage}
   \hspace{0.50cm}
\begin{minipage}{0.45\textwidth}
  \vspace{-0.10cm}
  \includegraphics[width=6.1cm]{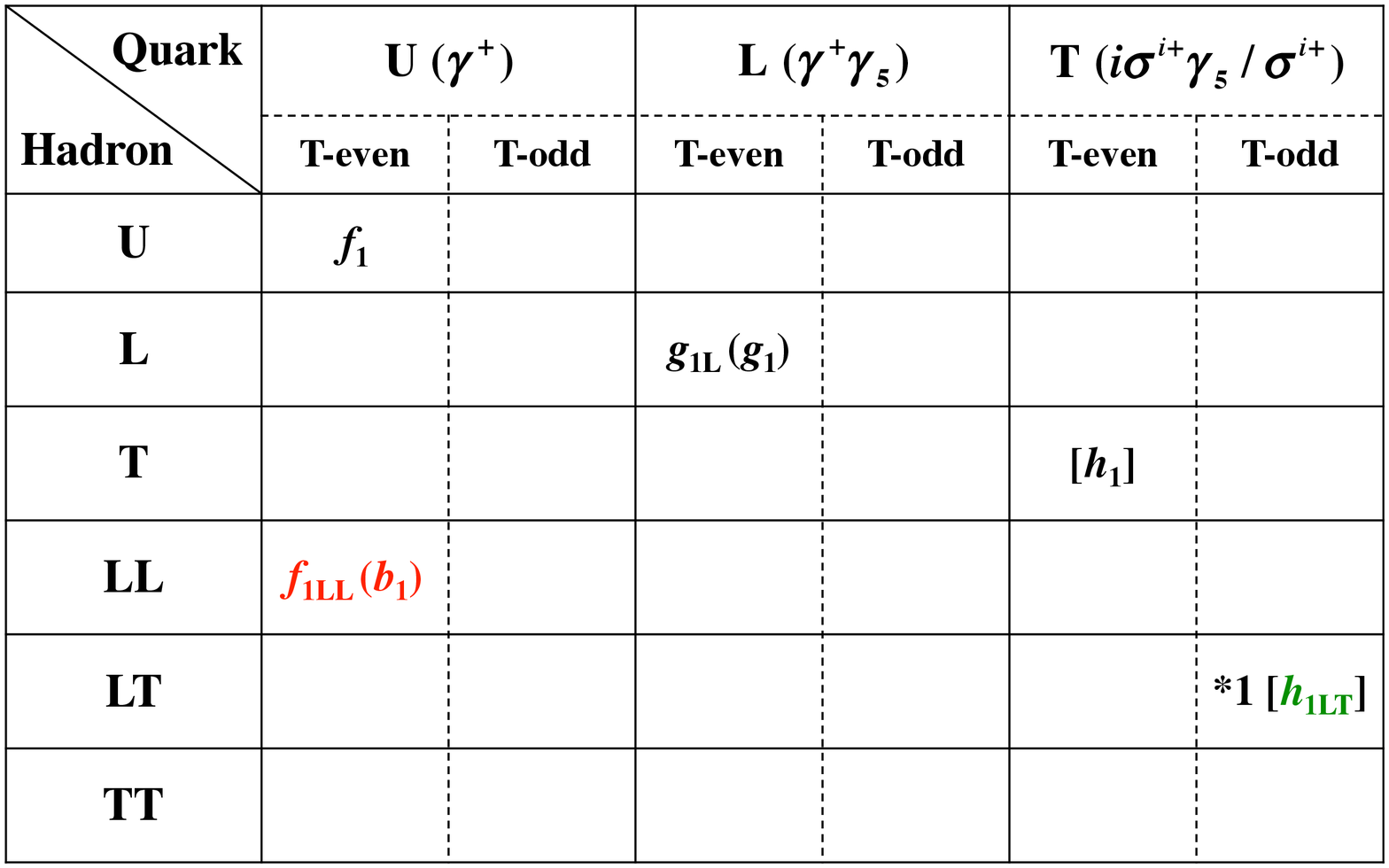}
  \vspace{-0.05cm}
  \caption{\hspace{-0.10cm} Twist-2 PDFs.}
  \vspace{-0.20cm}
  \label{table:twist-2-pdfs}
\end{minipage}
 \ \\[+0.05cm]
\begin{minipage}{0.45\textwidth}
  \includegraphics[width=6.4cm]{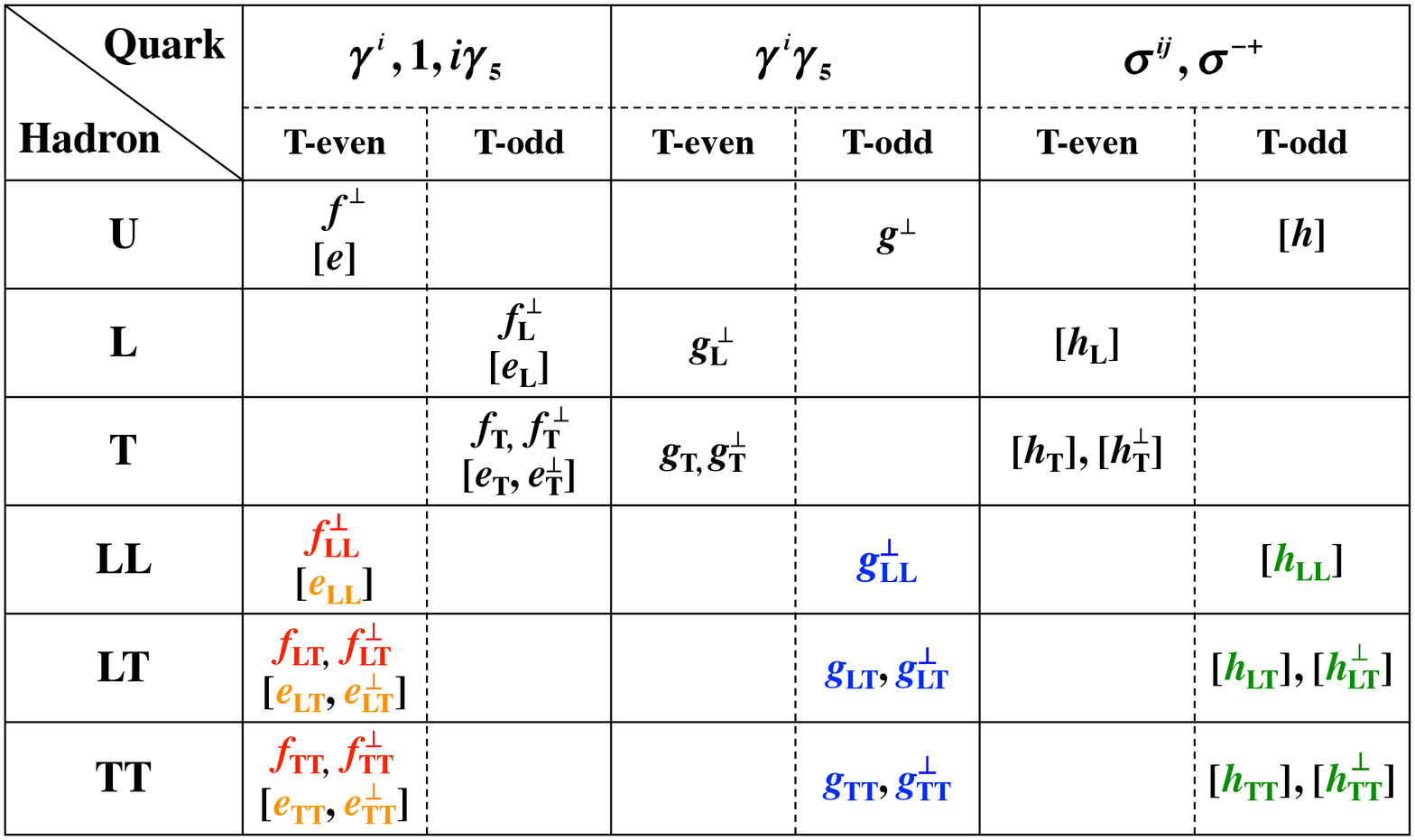}
  \vspace{-0.15cm}
  \caption{\hspace{-0.10cm} Twist-3 TMDs.}
  \label{table:twist-3-tmds}
\end{minipage}
    \hspace{0.50cm}
\begin{minipage}{0.45\textwidth}
  \includegraphics[width=6.1cm]{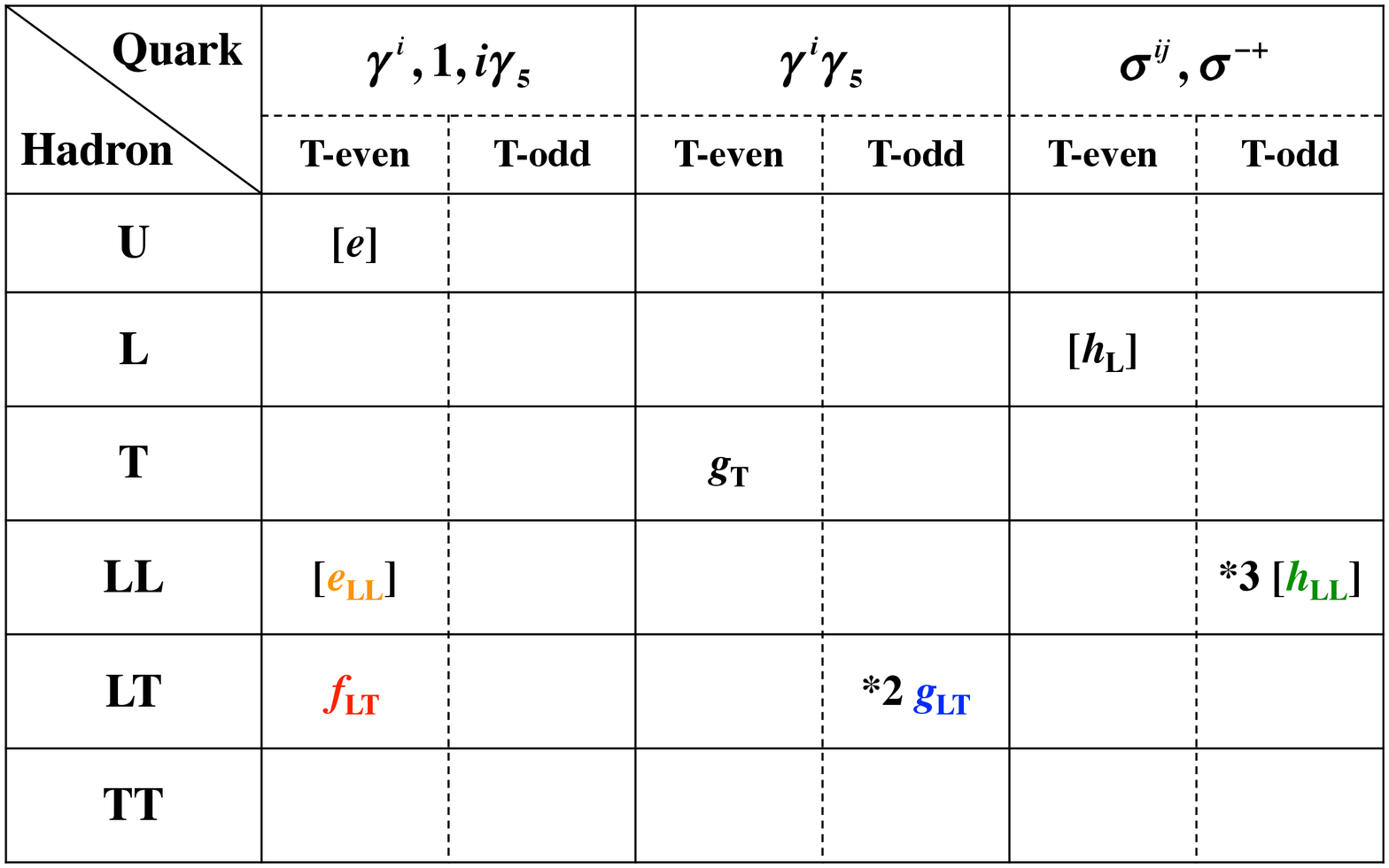}
  \vspace{-0.15cm}
  \caption{\hspace{-0.10cm} Twist-3 PDFs.}
  \label{table:twist-3-pdfs}
\end{minipage}
\ \\[0.05cm]
\begin{minipage}{0.45\textwidth}
  \includegraphics[width=6.4cm]{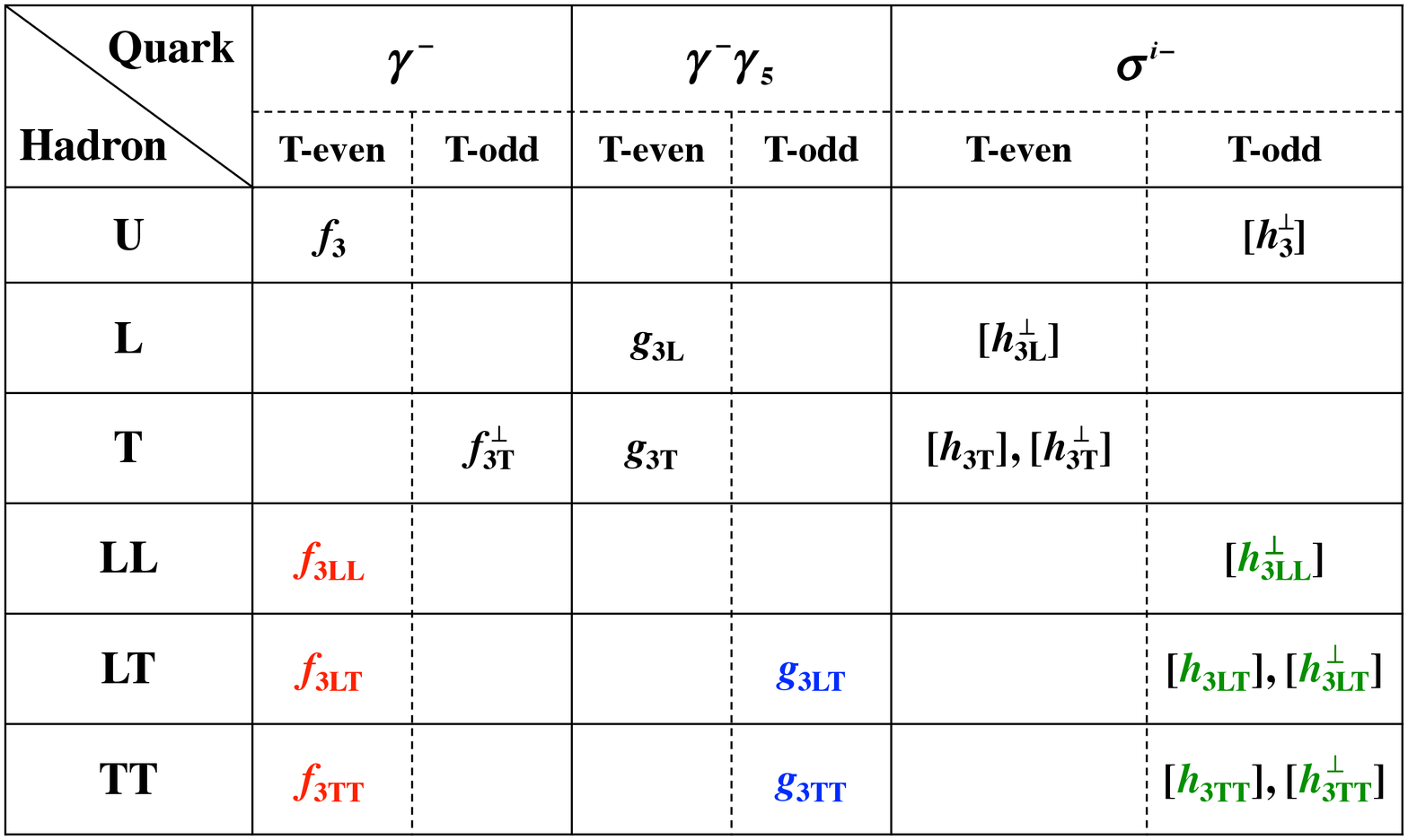}
  \vspace{-0.15cm}
  \caption{\hspace{-0.10cm} Twist-4 TMDs.}
  \label{table:twist-4-tmds}
\end{minipage}
    \hspace{0.50cm}
\begin{minipage}{0.45\textwidth}
  \includegraphics[width=6.1cm]{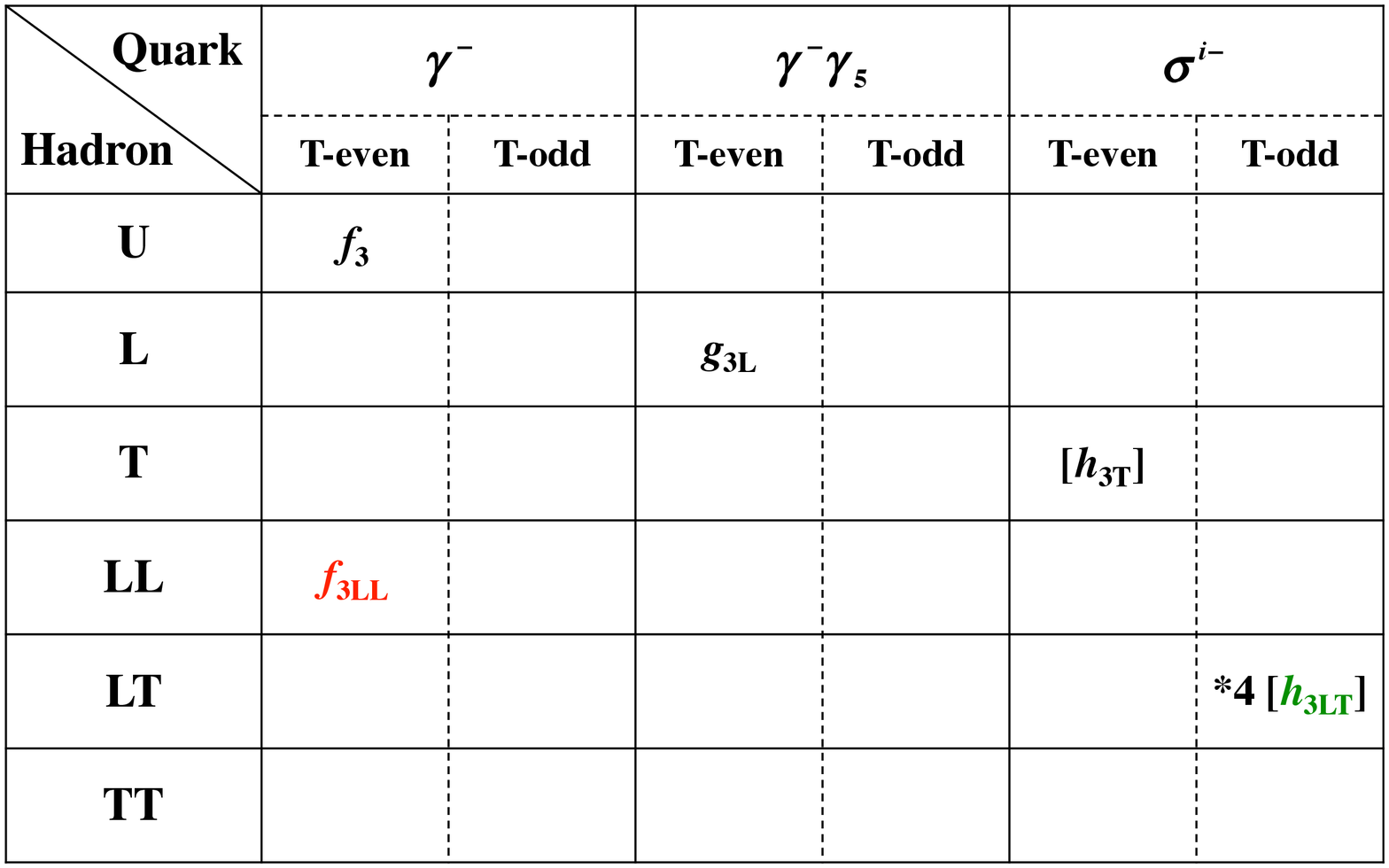}
  \vspace{-0.15cm}
  \caption{\hspace{-0.10cm} Twist-4 PDFs.}
  \label{table:twist-4-pdfs}
\end{minipage}
\end{table}

\section{TMD and collinear fragmentation functions 
         for a spin-1 hadron up to twist 4}

Fragmentation functions of spin-1 hadrons can be investigated
in the same way up to the twist-4. 
The collinear fragmentation functions were investigated up to the twist 4
in Ref.\,\cite{ji-ffs}; however, consistent studies on the TMDs 
of the spin-1 hadrons were restricted to the twist-2 level 
until recently.
Since we have obtained the twist-3 and 4 TMDs,
it became possible to investigate the TMD fragmentation functions
up to the twist 4. It is not necessary to investigate 
the TMD fragmentation functions from the beginning. 
Simply changing the kinematical variables and function names as follows
\vspace{-0.05cm}
\begin{align}
& \ \hspace{-0.20cm}
\text{Kinematical variables:}   \ \  
x, k_T, S, T, M, n, \gamma^+, \sigma^{i+}
\Rightarrow \ 
 z, k_T, S_h, T_h, M_h, \bar n, \gamma^-, \sigma^{i-},
\nonumber \\
& \ \hspace{-0.20cm}
\text{Distribution functions:}  \ \ f, g, h, e \hspace{2.30cm}
\ \Rightarrow \ 
\text{Fragmentation functions:} \ 
D, G, H, E ,
\label{eqn:tmd-fragmentation}
\nonumber \\[-0.60cm]
\end{align} 
we obtained the corresponding TMD fragmentation functions
\cite{ks-tmd-2021}.
The twist-2 TMD and collinear fragmentation functions are listed
in Tables \ref{table:twist-2-tmd-ffs} and \ref{table:twist-2-ffs}.
The twit-3 and twist-4 functions are given in Tables
\ref{table:twist-3-tmd-ffs}, \ref{table:twist-3-ffs},
\ref{table:twist-4-tmd-ffs}, and \ref{table:twist-4-ffs}.
We may note that the T-odd collinear fragmentation functions
exist, although the T-odd PDFs do not exist, because
the time-reversal invariance does not have to be imposed.
Since the spin-1 fragmentation functions can be measured,
for example, for the $\rho$ meson, these TMD and collinear
fragmentation functions are interesting quantities 
for future measurements.

\begin{table}[h]
\vspace{-0.00cm}
\begin{minipage}{0.48\textwidth}
  \includegraphics[width=6.2cm]{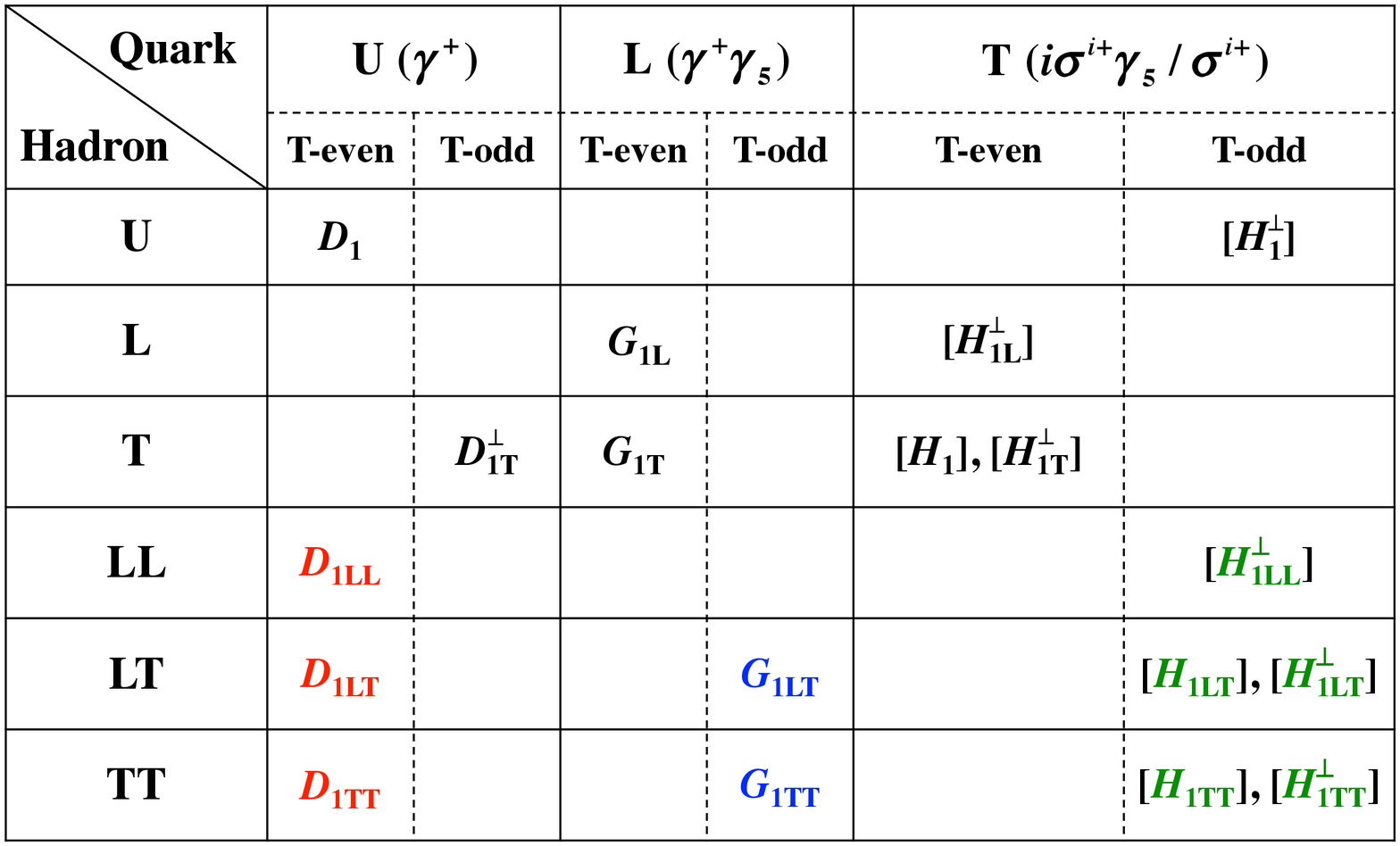}
  \vspace{-0.15cm}
  \caption{\hspace{-0.10cm} Twist-2 TMD fragmentation functions.}
  \label{table:twist-2-tmd-ffs}
\end{minipage}
   \hspace{0.35cm}
\begin{minipage}{0.47\textwidth}
  \vspace{0.10cm}
  \includegraphics[width=5.8cm]{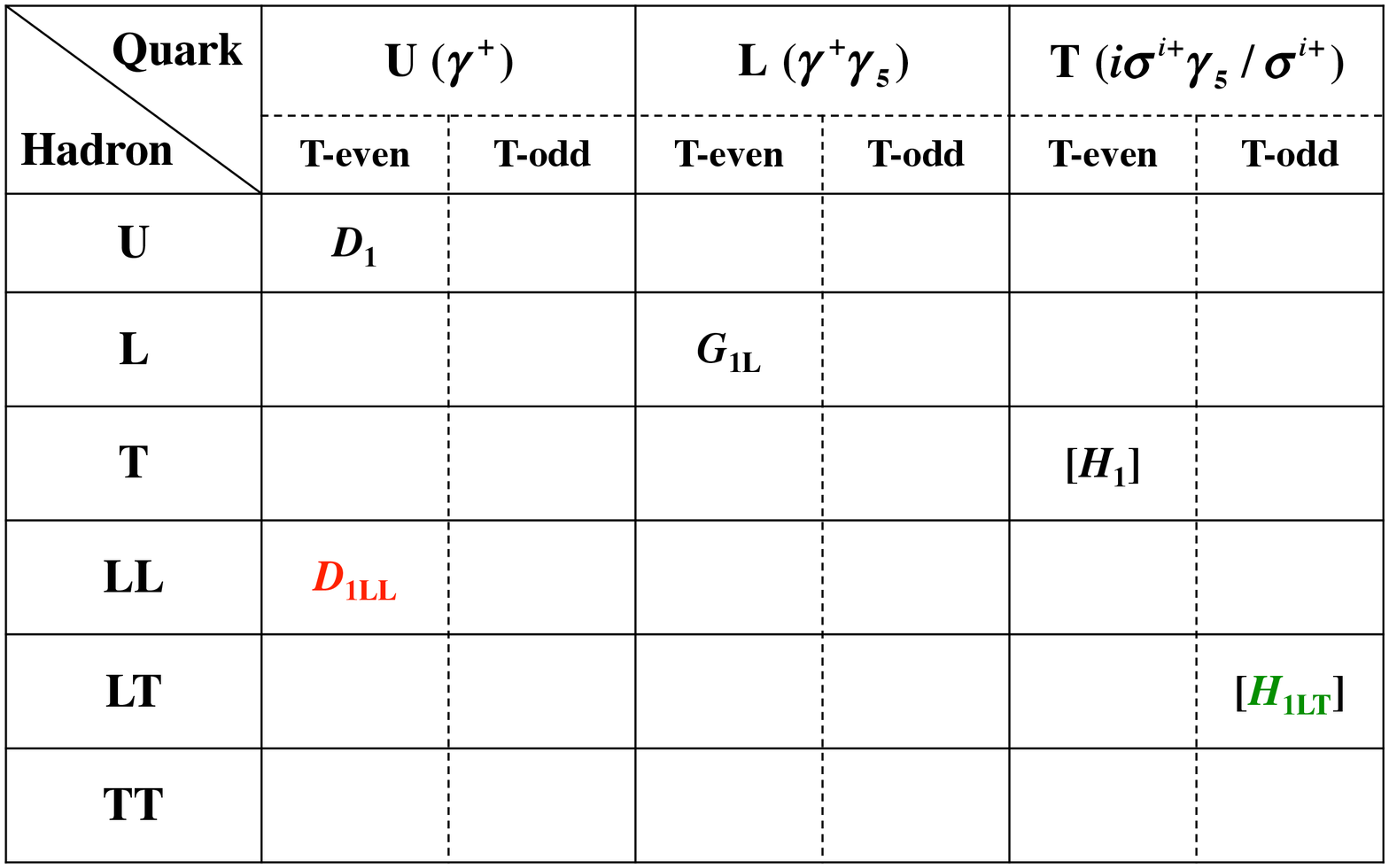}
  \vspace{-0.15cm}
  \caption{\hspace{-0.10cm} Twist-2 fragmentation functions.}
  \label{table:twist-2-ffs}
\end{minipage}
 \ \\[+0.15cm]
\begin{minipage}{0.47\textwidth}
  \includegraphics[width=6.4cm]{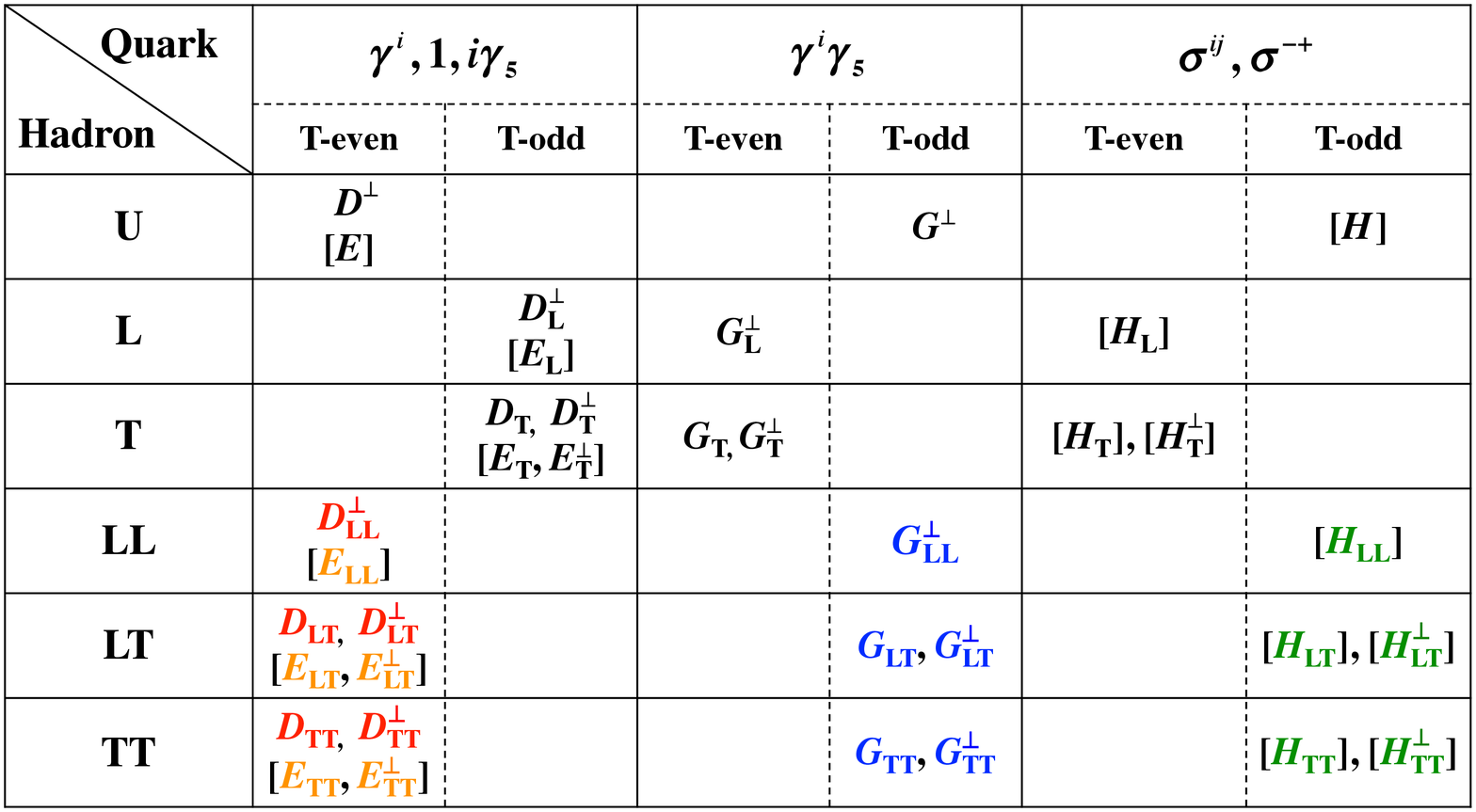}
  \vspace{-0.15cm}
  \caption{\hspace{-0.10cm} Twist-3 TMD fragmentation functions.}
  \label{table:twist-3-tmd-ffs}
\end{minipage}
    \hspace{0.50cm}
\begin{minipage}{0.47\textwidth}
  \includegraphics[width=5.8cm]{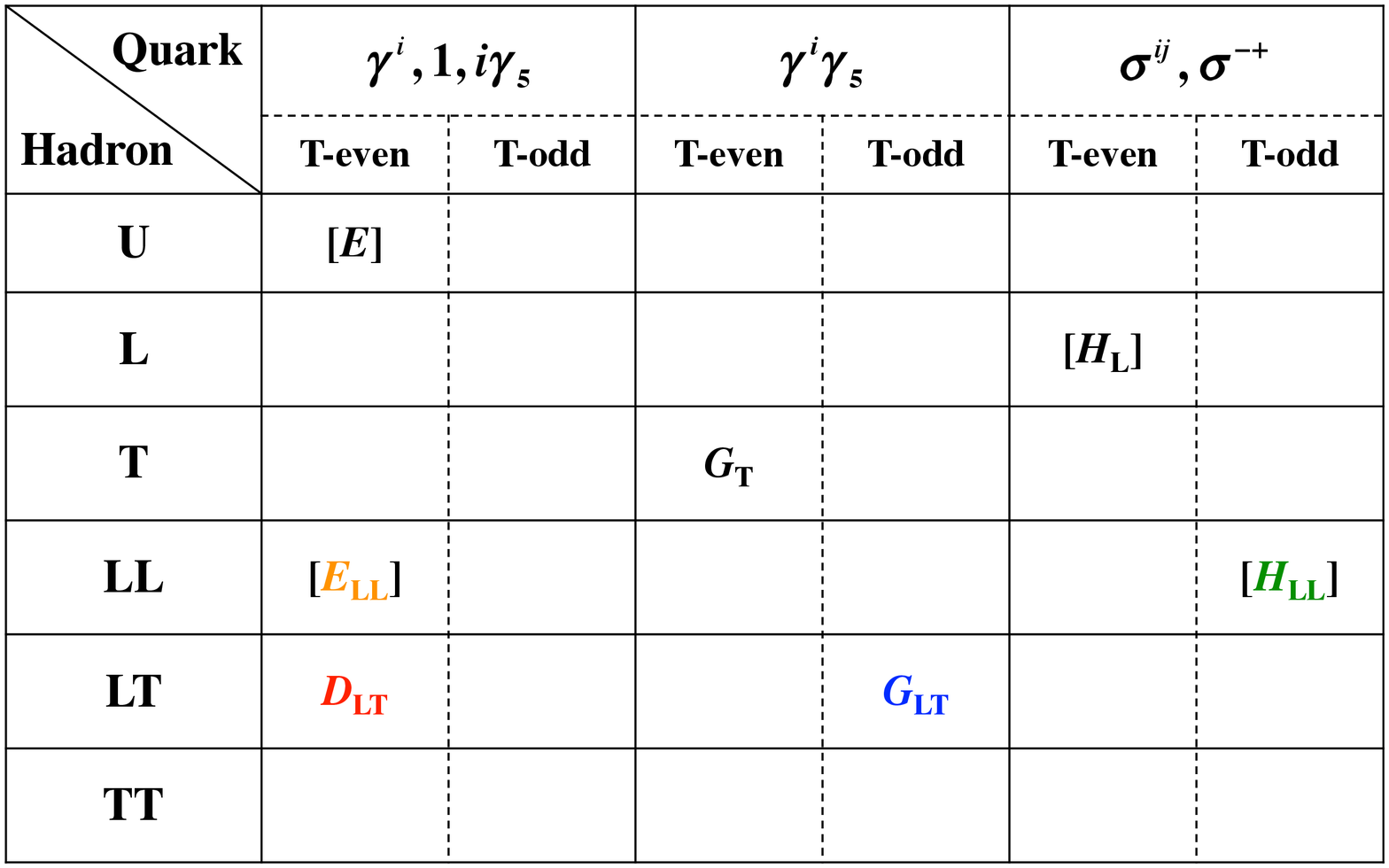}
  \vspace{-0.10cm}
  \caption{\hspace{-0.10cm} Twist-3 fragmentation functions.}
  \label{table:twist-3-ffs}
\end{minipage}
\ \\[0.15cm]
\begin{minipage}{0.47\textwidth}
  \includegraphics[width=6.4cm]{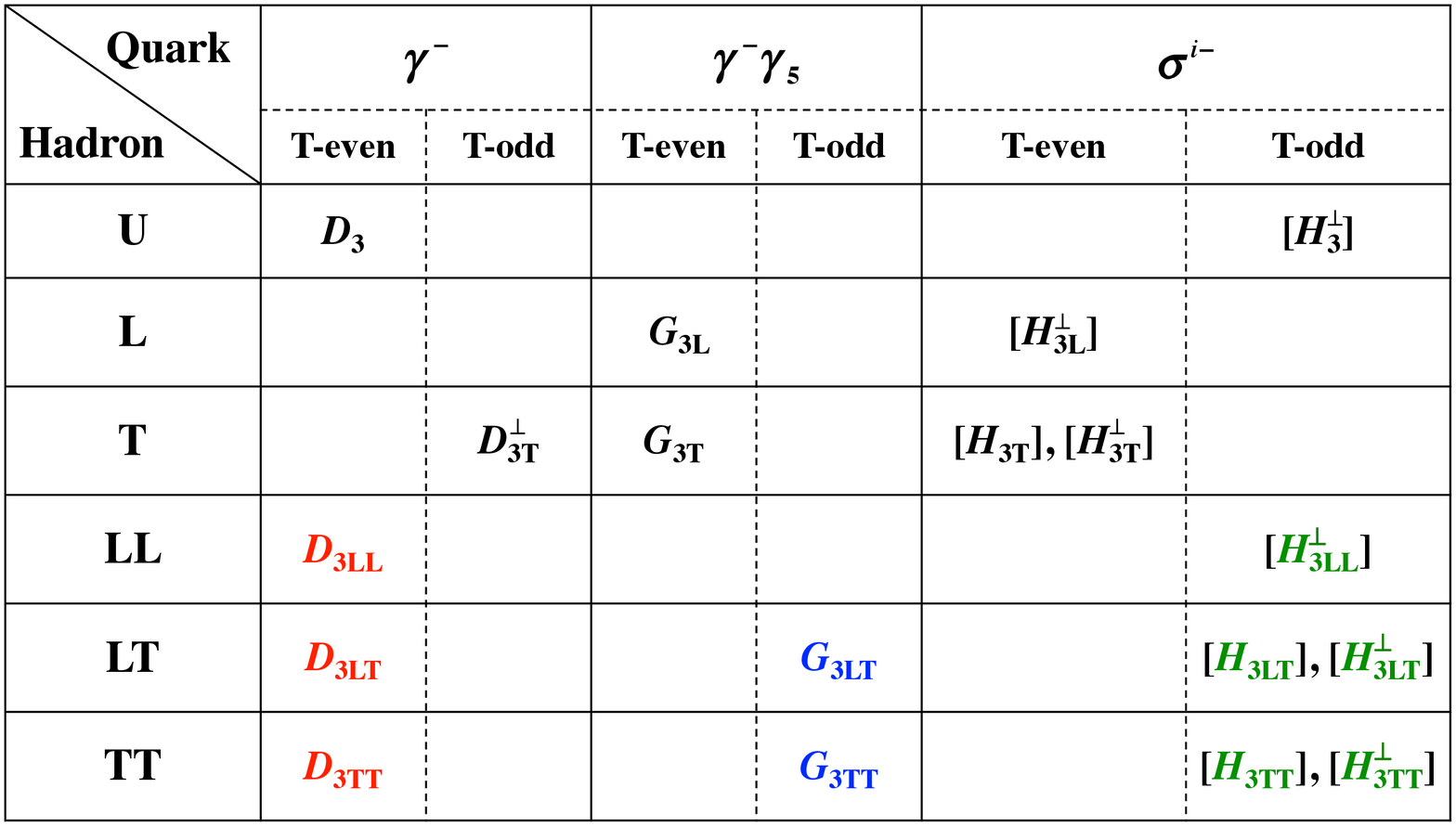}
  \vspace{-0.15cm}
  \caption{\hspace{-0.10cm} Twist-4 TMD fragmentation functions.}
  \label{table:twist-4-tmd-ffs}
\end{minipage}
    \hspace{0.50cm}
\begin{minipage}{0.47\textwidth}
  \includegraphics[width=5.8cm]{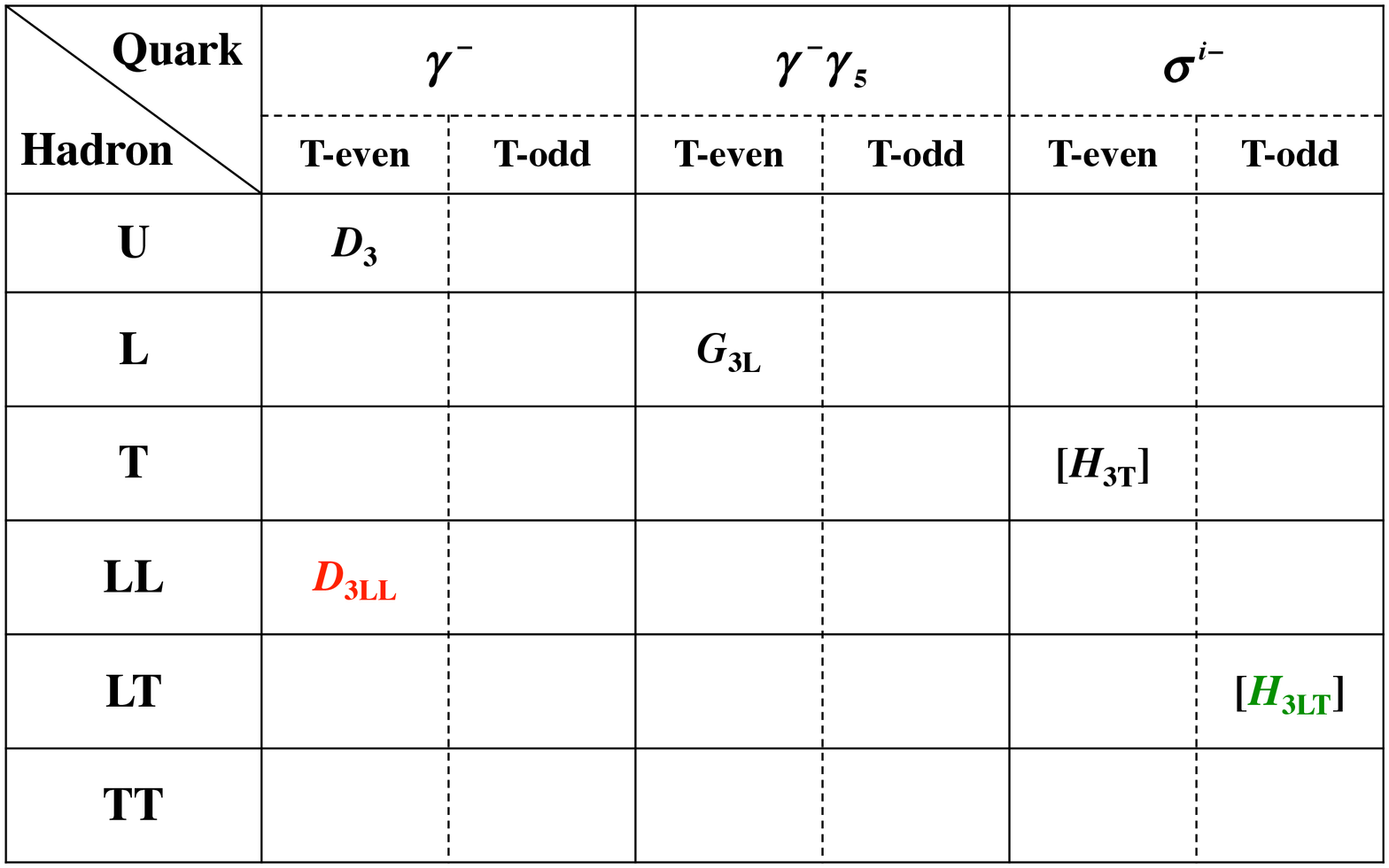}
  \vspace{-0.10cm}
  \caption{\hspace{-0.10cm} Twist-4 fragmentation functions.}
  \label{table:twist-4-ffs}
\end{minipage}
\end{table}

\section{Prospects and summary}

The TMD physics is one of popular hadron physics topics nowadays partly
because the TMD distributions are rare quantities to probe the explicit 
color flow within hadrons. They have been investigated for the nucleons so far.
However, since the polarized spin-1 deuteron targets and beams will be
available in the 2020's and 2030's at world 
accelerator facilities, we expect that the TMDs of the polarized deuteron
should become an interesting topic. Considering this situation, 
we showed all the possible TMDs, collinear PDFs, and fragmentation
functions up to the twist 4 in our theoretical studies. 
Now, it became possible to investigate them at the same level
with the nucleon structure functions including higher-twist effects
before the polarized-deuteron experiments. Since there are 
additional polarization quantities in the spin-1 deuteron, 
namely the tensor polarizations, our studies will 
become valuable along with future experimental investigations.

\appendix\section{}

S. Kumano was partially supported by 
Japan Society for the Promotion of Science (JSPS) Grants-in-Aid 
for Scientific Research (KAKENHI) Grant Number 19K03830.
Qin-Tao Song was supported by the National Natural Science Foundation 
of China under Grant Number 12005191, the Academic Improvement Project 
of Zhengzhou University, and the China Scholarship Council 
for visiting Ecole Polytechnique.




\begin{thebibliography}{9}
\bibitem{b1}
F. E. Close and S. Kumano, 
      {Phys. Rev. D {\bf 42}, 2377 (1990)};
   S. Kumano, 
      {Phys. Rev. D {\bf 82}, 017501 (2010)};
   S.~Kumano and Qin-Tao Song,
      {Phys. Rev. D {\bf 94}, 054022 (2016)};
   W. Cosyn, Yu-Bing Dong, S. Kumano, and M. Sargsian,
      {Phys. Rev. D {\bf 95}, 074036 (2017)}.    
\bibitem{g-transversity-drell-yan}
   S. Kumano, Qin-Tao Song,
      {Phys. Rev. D {\bf 101}, 054011 (2020)};
      {{\bf 101}, 094013 (2020)};
   S. Hino and S. Kumano, 
    {Phys. Rev. D {\bf 59}, 094026 (1999)};
    {{\bf 60}, 054018 (1999)}.
\bibitem{ks-tmd-2021}
   S. Kumano and Qin-Tao Song,
     {Phys. Rev. D {\bf 103}, 014025 (2021)}.
\bibitem{bm-2000} 
A. Bacchetta and P. Mulders,
     {Phys. Rev. D {\bf 62}, 114004 (2000)}.
\bibitem{tmds-nucleon}
  K.~Goeke, A.~Metz and M.~Schlegel,
     {Phys. Lett. B {\bf 618}, 90 (2005)};
  A.~Metz, P.~Schweitzer and T.~Teckentrup,
     {Phys. Lett. B {\bf 680}, 141 (2009)}.
\bibitem{ji-ffs} 
  X. Ji, 
     Phys. Rev. D {\bf 49}, 114 (1994).
\bibitem{eq-motion}
   S. Kumano and Qin-Tao Song, 
      {arXiv:2112.13218}, submitted for publication.
\bibitem{ks-ww-bc-2021}
   S. Kumano and Qin-Tao Song, 
      {JHEP {\bf 09}, 141 (2021)}.
\bibitem{ma-wang-zhang}
   J. P. Ma, C. Wang, and G. P. Zhang, 
      {arXiv:1306.6693} (unpublished, 2013).
\end{thebibliography}
\end{document}